\documentclass[final,1p,times]{elsarticle} 
\usepackage{graphicx} 
\usepackage{amssymb} 
\usepackage{amsthm} 
\journal{Nuclear Physics A} 
\begin{document} 

\begin{frontmatter} 


\title{Study of the crossover transition of a gas of extended hadrons}

\author{L.~Ferroni, V.~Koch}

\address{Nuclear Science Division, Lawrence Berkeley National Laboratory,
1 Cyclotron Road, Berkeley, 94720}

\begin{abstract} 
We formulate a simple model for a gas of extended hadrons at zero chemical potential by 
taking inspiration from the compressible bag model.
We show that a crossover transition qualitatively similar to lattice QCD can be 
reproduced by such a system by including some appropriate additional dynamics.
Under certain conditions, at high temperature, the system consists of
a finite number of infinitely extended bags, which occupy the entire space. In this situation
the system behaves as an ideal gas of quarks and gluons.
\end{abstract} 

\end{frontmatter} 




\section{The gas of compressible hadrons}\label{intro}
In this short note we will investigate a possible scenario for the crossover transition 
of a gas of hadrons at vanishing baryochemical potential 
(for a detailed description we suggest the reader to refer to~\cite{Ferroni:2008ej}).
Following the ideas proposed in~\cite{Gorenstein:1998am}, 
we adopt the philosophy that the same partition function should describe 
the system at both low and high temperature.
To this end we describe hadrons
as extended bags of Quark Gluon Plasma (QGP) in order to embody confined and 
deconfined phases from the very beginning. Accordingly (as derived in the simplest formulation 
of the MIT bag model~\cite{Chodos:1974je}), we start by assuming a hadron mass spectrum of 
the Hagedorn's type $\rho(m) \propto \exp(m/T_0)/m^\alpha$. 
To model the transition, however, we need to add some additional dynamics. 
We found that a qualitatively good agreement with lattice QCD (LQCD) can be obtained by simply 
taking into account the elastic interactions between hadrons together with excluded volume 
corrections. The elastic interactions give rise to a  {\em kinetic} 
pressure $p_k$ that in turn tends to ``squeeze'' the bag-like hadrons, resulting in a 
generalized temperature-dependent mass spectrum. 
The stability condition for the existence of a bag is given by the pressure balance:
\begin{equation}
p_r=B+p_k(V,T) \; ,
\label{simplass}
\end{equation} 
where $p_r$ is the internal pressure of the bag, B is the bag constant and $V$ and $T$ are the
system volume and temperature, respectively.
In our simple scheme, we adopt Boltzmann statistics and a non-relativistic framework, which is a good approximation 
in the vicinity of the transition region~\cite{Ferroni:2008ej}.  

In order to calculate the partition function of the system, one needs to evaluate 
the internal bags pressure $p_r$ in Eq.~(\ref{simplass}) that depends on the 
kinetic pressure $p_k$.
Because the pressure $p_k$ is thermally generated, it must be calculated 
from the partition function itself, resulting in a self-consistency relation.

\begin{figure}[h!]
\begin{center}
\includegraphics[width=0.7\textwidth]{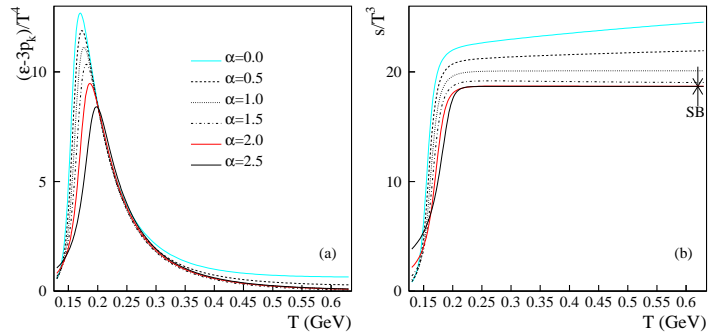}
\caption{\small{(Color online) Left panel (a): the ratio $(\varepsilon-3p_k)/T^4$.
Right panel (b): The ratio $s/T^3$, where $s$ is the entropy-density. }}
\label{isob2}
\end{center}
\end{figure}

The behavior of the system depends on the value of the 
parameter $\alpha$ of the mass spectrum.
For $\alpha > 5/2$ it exhibits a first order phase transition, whereas for $\alpha \leq 5/2$
the transition becomes a sharp crossover. 
Here, we analyze in detail the case of a crossover transition, i.e. $\alpha \leq 5/2$. 
As an example we show the ratios $(\varepsilon-3p_k)/T^4$
(Fig.~(\ref{isob2}a)) and $s/T^3\equiv (\varepsilon+p_k)/T^4$ (Fig.~(\ref{isob2}b)), 
where $\varepsilon$ and $s$ are the energy and the entropy density, respectively.
As one can see from (Fig.~(\ref{isob2}b)), for $\alpha=0$ and $1/2$, 
the curves grow with the temperature with larger slopes for smaller $\alpha$'s.
Instead, for $1 \leq \alpha \leq 5/2$, they settle onto constant asymptotic values 
in a qualitatively good agreement with LQCD.
As $\alpha$ changes from $1$ to $5/2$ the asymptotic value converges very fast to 
the Stefan-Boltzmann limit from above~\footnote{In this work
we ignored the $\sim 10 \%$ deviation of the LQCD result 
from the free gas (Stefan-Boltzmann) limit~\cite{Cheng:2007jq}.}. 
In ref.~\cite{Ferroni:2008ej}, we also studied
the ratios $p_k(\infty, T)/T^4$ and $\varepsilon/T^4$. Also these quantities exhibit a
similar behavior. In the range $1 \leq \alpha \leq 5/2$, the model seems to produce a smooth 
crossover transition 
toward a new regime whose features are very similar to those of a gas of massless particles, 
even though no deconfined states are included in the partition function. 
We have also studied the (strong) dependence of the 
particle density $\langle n \rangle = \langle N \rangle/V$ (where $V \rightarrow \infty$) 
on $\alpha$. We have found that there exist a limiting value $\alpha_0$ between $2.12$ and $2.13$ 
such that for $\alpha_0<\alpha \leq5/2$ the particle density vanishes at high temperature. 
The system is then populated by one (or few) infinite bag(s) that occupies the entire volume.
Conversely, for $\alpha < \alpha_0$, $\langle n \rangle$ grows with the temperature. 
In the range $1 \leq \alpha < \alpha_0$, the ideal gas behavior is mimicked
by a number of heavy extended bags that saturate the phase space forming a dense system.


\section*{Acknowledgments}
This work is supported  by the Director, Office of Energy
Research, Office of High Energy and Nuclear Physics, Divisions of
Nuclear Physics, of the U.S. Department of Energy under Contract No.
DE-AC02-05CH11231.
\end{document}